\documentclass[11pt]{article}
\newcommand{\capstyle}{\footnotesize}
\usepackage{graphics}
\input epsf
\epsfverbosetrue
\def\styleversion{1.002}\def\styledate{7 Jul 1995}
\def\titlepage#1#2{\clearpage%
\setcounter{footnote}{0}\pagestyle{empty}%
\mbox{\begin{tabular}{llr}
\epsfxsize=20mm
\epsfbox{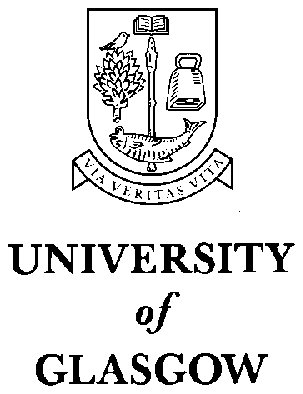}
& \begin{tabular}{c} 
\vspace*{-2.7cm}\\
{\Large Department of Physics \& Astronomy} \\
\multicolumn{1}{c}{\large Experimental Particle Physics Group} \\
{\small Kelvin Building, University of Glasgow,} \\
{\small Glasgow, G12 8QQ, Scotland} \\
{\small Telephone: +44 (0)141 339 8855 Fax: +44 (0)141 334 9029}\\
\end{tabular} 
& \begin{tabular}{r}
\vspace*{-5cm}\\
{\bf #1} \\
{#2}\\
\end{tabular}
\end{tabular}}

\vspace{2cm}

}% End of \titlepage tag
\def\collaboration#1{\vskip1em\begin{center}\it#1\end{center}}
\def\conference#1{\vskip1em\begin{center}\it#1\end{center}}
\def\title#1{\vskip1em\begin{center}\large\bf#1\end{center}\vskip1.5em}
\def\abstract{\begin{center}{\bf Abstract}\\[\baselineskip]%
\end{center}\quotation\small}

\def\endtitlepage{%% Reset counters
\setcounter{footnote}{0}\let\titlepage\relax\vfill
\newpage\setcounter{page}{1}\pagestyle{plain}\pagenumbering{arabic}%
\gdef\@thanks{}\gdef\@author{}\gdef\@title{}\let\thanks\relax}
\begin{document}
\bibliographystyle{mine}
\begin{titlepage}{GLAS--PPE/96--XX}{\today}
\title{Rapidity Gaps in Hard Photoproduction}
\centerline{L.E. Sinclair\footnote{e-mail: sinclair@desy.de}}
\collaboration{for the ZEUS Collaboration}
\conference{Talk presented at the Topical Conference on Hard Diffractive 
Processes,\\ Eilat, Israel, February 1996.}

\begin{abstract}
Recent results obtained from studies of diffractive processes in hard 
photoproduction performed by the ZEUS collaboration using data delivered
by HERA in 1993 and 1994 are presented.
In particular, we have found that $(7 \pm 3)$\% of events with two jets at a
pseudorapidity interval of 3.5 to 4 are inconsistent with a non-diffractive
production mechanism.  These events may be interpreted as arising due to
the exchange of a colour singlet object of negative squared invariant 
mass ($-t$) around
40~GeV$^2$.  We have also probed the structure of the exchanged colour singlet
object in low--$t$ diffractive
scattering.  By comparing the results from photoproduction and
electroproduction processes we find that between 30\% and 80\% of the momentum
of the exchanged colour singlet object which is carried by partons
is due to hard gluons.
\end{abstract}
\end{titlepage}
%%%%%%%%

\section{Introduction}

In this first section a brief introduction to hard photoproduction is
presented.  Then the general characteristics of the photoproduction events 
which give rise to rapidity gaps in the final state are described and
diffraction is defined in this context.  The events may be classified
into two groups, those which give rise to a central rapidity gap, and those
which give rise to a forward rapidity gap.  The results which have been
obtained by the ZEUS Collaboration from the study of these two classes
of events are presented and discussed in the following two sections.  These 
are published results~\cite{Zrgbj,Zlowt3}, and the reader is referred to the
publications for detailed accounts of the event selection,
the Monte Carlo event generation and the corrections for detector effects.
Some concluding remarks and an outlook are provided in the final section.

\subsection{Hard photoproduction}
\begin{figure}[h]
\epsfxsize=12.5cm
\centering
\leavevmode
\epsfbox{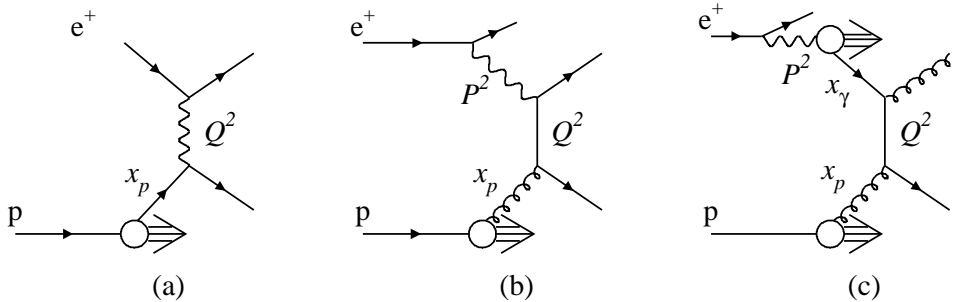}
\caption{\capstyle\label{f:hphp}
        Diagrams showing HERA processes.  The canonical electroproduction
        process is shown in (a).  A leading order direct photoproduction
        process is shown in (b) while an example of a leading order
        resolved photoproduction process is shown in (c).}
\end{figure}

The canonical HERA event proceeds as illustrated in Figure~\ref{f:hphp}(a).  
The incoming positron is scattered through a large angle exchanging a 
photon probe of (negative) virtuality as high as 
$Q^2 \sim 5 \cdot 10^4$~GeV$^2$.  
The structure of the proton may be studied down to values of the 
Bjorken--$x_p$ variable as low as $x_p \sim 5 \cdot 10^{-3}$.
Of course the electroproduction cross section is strongly peaked to 
$Q^2 \sim 0$ and the events most copiously produced at HERA are soft 
photoproduction events.  However photoproduction events which lead to the
production of high transverse energy jets in the final state are also 
characterized by a large (negative) squared momentum transfer $Q^2$.
An example is shown in 
Figure~\ref{f:hphp}(b).  For these hard photoproduction events the negative
of the squared invariant mass of the photon is denoted $P^2$ and of course,
$P^2 \sim 0$.  Again, very low values of $x_p$ of the proton may be
probed and note that the photoproduction processes (in contrast to the
electroproduction processes) are directly sensitive to the gluon content
of the proton.

The incoming photon may fluctuate into a hadronic state
before interaction with the proton.  This situation is illustrated in
Figure~\ref{f:hphp}(c).  The momentum fraction variable $x_{\gamma}$ has
been introduced, where $x_{\gamma}$ represents the fraction of the photon's 
momentum which participates in the hard interaction.  The class of events
represented by Figure~\ref{f:hphp}(b) are known as direct photoproduction
events and have $x_{\gamma} = 1$.  Resolved photoproduction events are 
represented by 
Figure~\ref{f:hphp}(c) and have $x_{\gamma} < 1$.  The present discussion is
clearly limited to leading order processes although a definition of
$x_{\gamma}$ may be made which is calculable to all orders and allows for
a well defined separation of direct and resolved photoproduction 
processes~\cite{Zdij93}.

\begin{figure}[h]
\epsfxsize=8cm
\centering
\leavevmode
\rotatebox{270}{\epsfbox{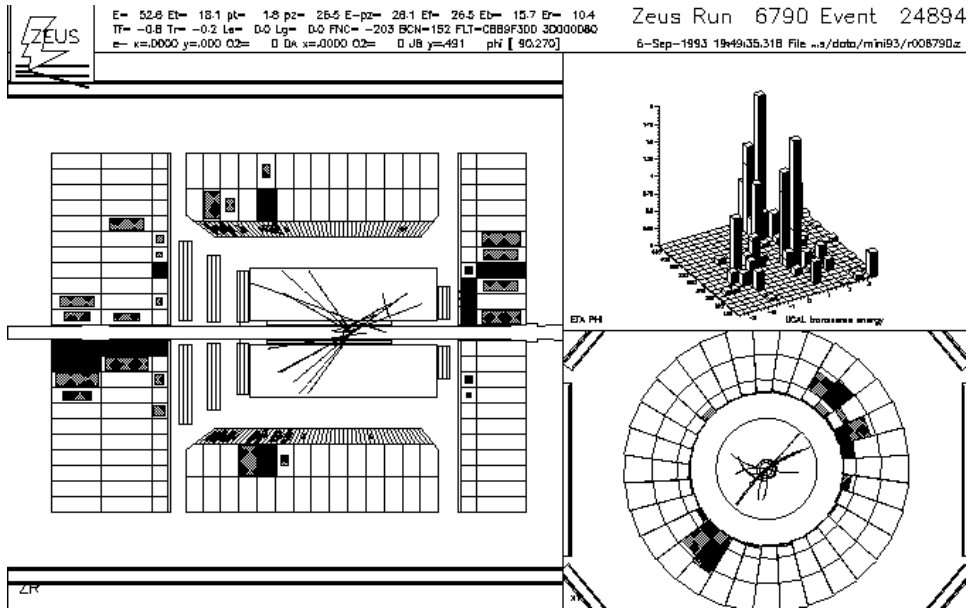}}
\caption{\capstyle\label{f:hphpevt}
        A hard photoproduction event in the ZEUS detector.  The $z -R$
longitudinal view is shown on the left hand side.  In the upper right hand 
corner the $\eta$ and $\varphi$ coordinates of the hit calorimeter cells are 
shown, weighted by their transverse energies.  In the lower right hand corner
the $x - y$ or transverse view is shown.}
\end{figure}

A hard photoproduction event in the ZEUS detector is shown in 
Figure~\ref{f:hphpevt}.  In the $z - R$ display on the left--hand side the
positrons approach from the left and the protons from the right.  The 
$e^+$ beam has an energy of 27.5~GeV and the $p$ beam has an energy of 
820~GeV.  The calorimeter is deeper in the ``forward'' or proton direction,
to cope with this asymmetry in the beam energies.
This proton direction is the direction of positive pseudorapidity,
$\eta = -\ln \tan (\vartheta / 2)$, where $\vartheta$ is the polar angle with
respect to the $p$ beam direction.
Two jets of large transverse energy are measured in the tracking chambers
and the calorimeter and are clearly apparent in all three views.
The jets are both at $\eta \sim 1$ ($\vartheta \sim 40^{\circ}$)
and are back to back in $\varphi$.  It
is the energy deposits and tracks of these jets which we use to select a 
sample of hard photoproduction events.  Notice that there is
a large energy deposit in the far--forward region next to the beam pipe.  This
energy is associated with the proton remnant.  There is also a large energy
deposit in the rear direction which could be called the photon remnant if 
this were considered a resolved photon event.  (The transverse energy of the
rear jet in this particular event is actually sufficiently large that it 
may be appropriate to consider this a higher order direct photoproduction
event.)  Notice that there is no energy deposit which could be associated 
with the scattered $e^+$, which is lost down the rear beam pipe in 
photoproduction processes.

\subsection{Diffraction\label{s:DIFF}}
The analyses which will be discussed in this report both make use of the 
operational definition of diffraction~\cite{bj2}:
\begin{quote}\label{g:diff}
   {\em A process is diffractive if and only if there is a large
   rapidity gap in the produced--particle phase space which is not
   exponentially suppressed.}
\end{quote}
They are, in addition, studies of {\em hard diffraction} in the sense that the
events all possess a large (negative) squared momentum transfer, $Q^2$, or
a high energy scale, $Q$.  The hard diffraction 
events are further subdivided into two classes both of which have gone by
a number of different names.

\begin{figure}[h]
\epsfxsize=7cm
\centering
\leavevmode
\epsfbox{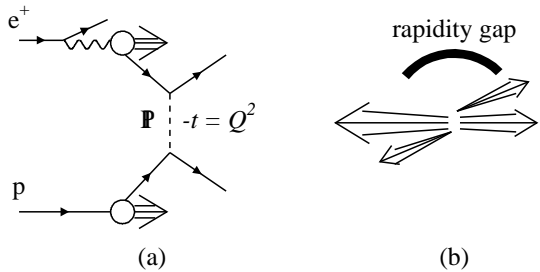}
\caption{\capstyle\label{f:hight}
        Hard diffractive scattering at HERA.  The diagram for this
        process is shown
        in (a).  The exchanged colour singlet object is denoted $I\!\!P$ and
        the negative of its squared invariant mass, $-t$, sets the energy
        scale of the interaction, ($Q = \protect\sqrt{-t}$).
        In the final state, shown in (b),
        there are two high transverse energy jets and two remnant jets with
        a gap in particle production in the central rapidity region.}
\end{figure}

The first class of events may be called
hard diffractive scattering, hard double--dissociation diffraction or 
high--$t$ diffraction.
They proceed as shown in Figure~\ref{f:hight}(a), via the exchange of a 
colour singlet object of large negative squared invariant mass, $t$.
($t$, in both event classes,
refers to the square of the momentum transfer across the exchanged colour 
singlet
object.  This object is called a pomeron and denoted $I\!\!P$.)
Owing to the absence of colour flow across the middle of the event a gap in 
the production of particles is expected to be observable.
These events thus contain a central rapidity gap as illustrated in 
Figure~\ref{f:hight}(b).
This may be contrasted with the situation, for example, where a gluon is 
exchanged in place of the pomeron in Figure~\ref{f:hight}(a).
Central rapidity gap events will be examined in Sect.~\ref{s:CRG}.

\begin{figure}[h]
\epsfxsize=7cm
\centering
\leavevmode
\epsfbox{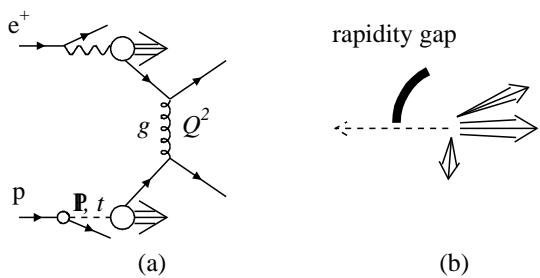}
\caption{\capstyle\label{f:lowt}
        The diffractive hard photoproduction process at HERA is shown in (a).
        The pomeron, denoted $I\!\!P$, is shown being emitted from the
        proton with a squared momentum transfer $t$.  A quark from the 
        pomeron 
        subsequently enters the hard
        subprocess which is mediated by the exchange of a gluon, 
        denoted $g$, and
        characterized by the energy scale $Q$.  The
        topology of the final state is shown in (b).  There are two high
        transverse energy jets associated with the hard subprocess.  There
        may be a photon remnant.  However the proton is not broken up and
        disappears down the forward beam pipe leaving a gap in particle
        production at high rapidities.}
\end{figure}

The second class of events has been called diffractive hard scattering,
hard single--dissociation diffraction and low--$t$ diffraction.  These events 
are understood to occur when a colour singlet object, travelling collinearly
with the proton, is probed by the hard subprocess.  An example is
shown in Figure~\ref{f:lowt}(a).  
Because the object emitted by the proton
does not carry colour, particle production into the forward, or high--$\eta$,
region of phase space is suppressed.  This process thus leads to the
formation of a forward rapidity gap as illustrated in 
Figure~\ref{f:lowt}(b).  This process is studied in Sect.~\ref{s:FRG}.

\section{Central Rapidity Gaps\label{s:CRG}}

The results discussed in this section have been published in~\cite{Zrgbj}.
We have isolated a sample of hard photoproduction events containing at 
least two jets of transverse energy $E_T^{jet} > 6$~GeV.  The jets are
found using a cone algorithm with jet cones of radius 1 in 
$\eta - \varphi$ space.  The pseudorapidity interval between the jet centres,
$\Delta\eta$, exceeds 3.5 in 535 of the 8393 events.
Note that to leading order $\Delta\eta = \ln (\hat{s} / -\hat{t})$ where
$\hat{s}$ and $\hat{t}$ are the usual Mandelstam variables of the hard
subprocesses.  $\Delta\eta > 3.5$ therefore means that 
$\hat{s} > 30 \cdot - \hat{t}$ which falls into the Regge regime, 
$\hat{s} \gg - \hat{t}$.

Gap candidate events are
defined as those which have no particles of $E_T^{particle} > 300$~MeV
between the edges of the jet cones in pseudorapidity.  
The size of the gap therefore lies between $\Delta\eta$ and 
$\Delta\eta -2R = \Delta\eta -2$.

\begin{figure}[h!]
\epsfxsize=8cm
\centering
\leavevmode
\rotatebox{270}{\epsfbox{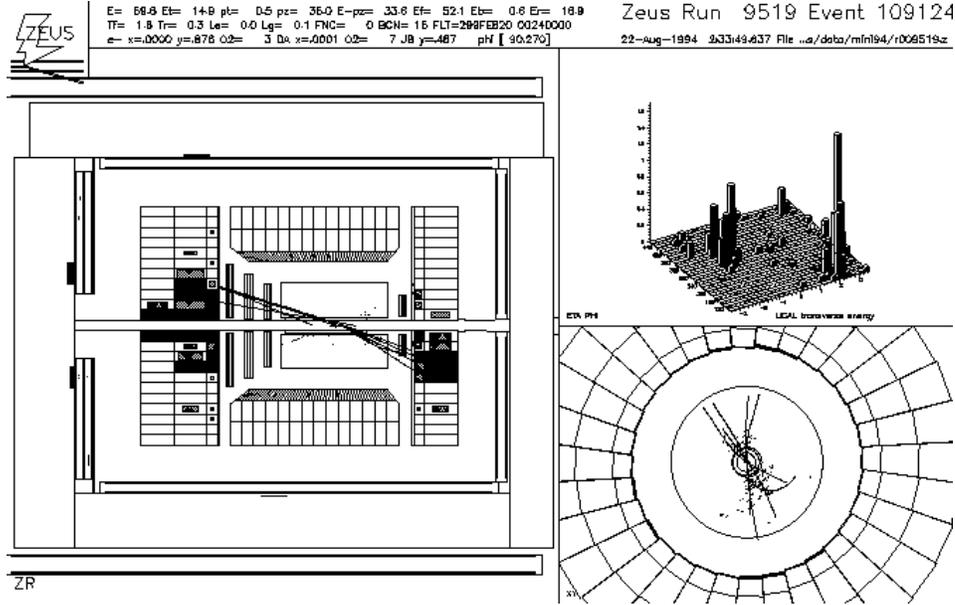}}
\caption{\capstyle\label{f:cgapevt}
        A hard photoproduction event with a central gap in the ZEUS
        detector.  The $z-R$ view of the ZEUS detector is shown on
        the left side.  The lego plot of the $E_T$ weighted energy 
        deposits in the calorimeter versus their $\eta$ and $\varphi$ is
        shown in the upper right picture and the lower right picture
        shows the $x-y$ cross section through the ZEUS detector.}
\end{figure}

An event from this sample is shown in Figure~\ref{f:cgapevt}.
There are two high transverse energy jets in this event which are back
to back in $\varphi$ and have a pseudorapidity interval of $\Delta\eta = 3.6$.
There are additional energy deposits around the forward beam pipe which 
correspond to the proton remnant and energy deposits near the rear beam pipe 
which may be associated with the photon remnant.
This is in fact a gap candidate event.  There are no candidate particles in 
the pseudrapidity interval between the jet cones having a transverse energy
of $E_T^{particle} > 300$~MeV.  There are, however, some very low energy 
energy deposits in this region which could in some cases be due to 
calorimeter noise.
Alternatively they may be particles which are so soft that they have no 
memory of their parent parton's direction.  The $E_T^{particle}$ threshold
is a necessary theoretical tool~\cite{muellertang,delduca,pumplin}
as well as a convenient experimental cut.

\begin{figure}[h!]
\epsfxsize=12cm
\centering
\leavevmode
\epsfbox{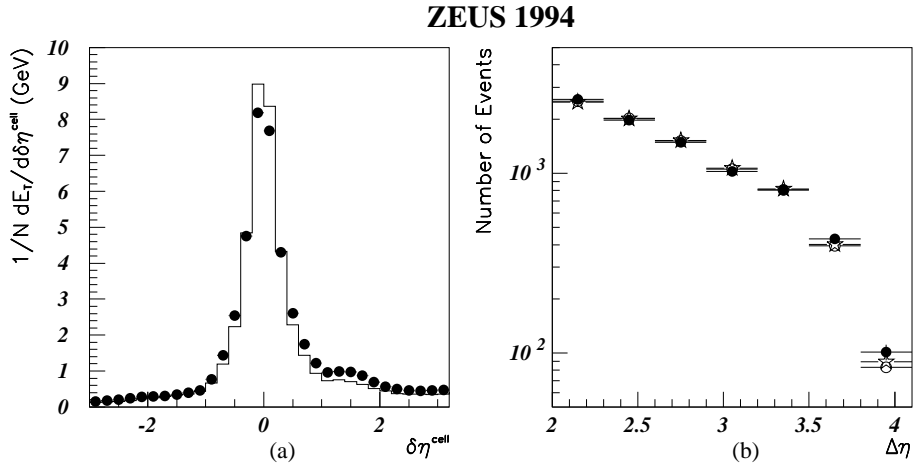}
\caption{\capstyle\label{f:cgap1}
        Sample characteristics.  Errors are statistical only.  No correction
        for detector effects has been performed.  Monte Carlo simulated events
        have been subjected to full detector simulation.
        (a) Jet profile.  Data are shown as black dots and PYTHIA standard
        hard photoproduction processes are shown as the solid line.
        (b) The $\Delta\eta$ distribution.  Data are shown as black dots, 
        the PYTHIA standard sample is shown by the open circles and the
        PYTHIA sample containing 10\% of photon exchange processes is
        shown by the stars.}
\end{figure}

The characteristics of this event sample are illustrated in 
Figure~\ref{f:cgap1}.  Here the data are shown uncorrected for any detector
effects, as black dots.  The errors shown are statistical only.  
The data are compared
to predictions from the PYTHIA~\cite{PYT1,PYT2} generator for hard 
photoproduction processes.  These predictions have been passed through a 
detailed simulation of the selection criteria and of the detector acceptance 
and smearing.

Figure~\ref{f:cgap1}(a) shows the average profile of the two highest 
$E_T^{jet}$
jets.  In the jet profile, $\delta\eta^{cell} = \eta^{cell} - \eta^{jet}$ of
each calorimeter cell is plotted, weighted by the cell transverse energy,
for cells with $|\varphi^{cell} - \varphi^{jet}|$ less than one radian.
The data show good collimation and a jet pedestal which increases gradually
towards the forward direction.
The PYTHIA prediction
for the standard direct and resolved hard photoproduction processes is shown 
by the solid line for comparison.  The description is reasonable, however
there is a slight
overestimation of the amount of energy in the jet core and
underestimation of the jet pedestal.  Higher order processes and secondary
interactions between photon and proton spectator particles are neglected in
this Monte Carlo simulation.  It is anticipated that their inclusion could
bring the prediction into agreement with the data~\cite{H1MIE,ZMIE}.  
Notice that, na\"{\i}vely, this
discrepancy would be expected to give rise to proportionally fewer events
containing a rapidity gap in the data than in the Monte Carlo sample.

Figure~\ref{f:cgap1}(b) shows the magnitude of the pseudorapidity interval 
between the two highest transverse energy jets, $\Delta\eta$.  The number
of events is rapidly falling with $\Delta\eta$ but we still have a sizeable
sample of events with a 
large value of $\Delta\eta$.
This distribution is well described by the standard PYTHIA simulation of
photoproduction events which is here represented by open circles.  The stars
show a special PYTHIA sample which has been introduced in this analysis 
primarily for the purpose of obtaining a good description of the data and
understanding detector effects.  90\% of this sample is due to the standard
photoproduction processes.  The other 10\% of this sample is due to 
quark--quark 
scattering via photon exchange (Figure~\ref{f:hight}(a) with
the $I\!\!P$ replaced by a $\gamma$) and obviously this 10\% contains no 
contribution from leading order direct photoproduction processes.
(Note that 10\% is about two orders of magnitude higher than one would
obtain from the ratio of the electroweak to QCD cross sections.)
The combined Monte Carlo sample also
provides a good description of the $\Delta\eta$ distribution.

\begin{figure}[h]
\epsfxsize=6cm
\centering
\leavevmode
\epsfbox{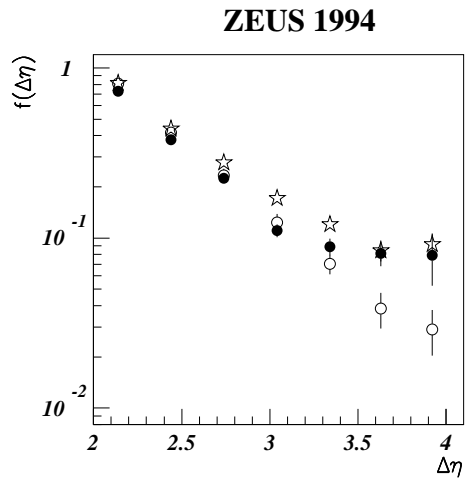}
\caption{\capstyle\label{f:cgap2}
        The distribution of the fraction of events containing a gap, 
        $f(\Delta\eta)$, with respect to $\Delta\eta$.  The black dots
        represent the data, the open circles represent the standard
        hard photoproduction simulated events and the stars represent
        the simulated event sample of which 10\% is due to photon
        exchange processes.  The errors  
        are statistical only, no correction
        for detector effects has been made, and the Monte Carlo samples have
        been passed through a detailed simulation of the ZEUS detector
        acceptance and smearing.}
\end{figure}

We define the gap--fraction, $f(\Delta\eta)$, to be the
fraction of dijet events which have no particle of 
$E_T^{particle} > 300$~MeV in 
the rapidity interval between the edges of the two jet cones.
The gap--fraction, uncorrected for detector effects, is shown in
Figure~\ref{f:cgap2}.  The data are shown as black dots, the events from
the standard PYTHIA simulation are shown as open circles and the events from
the PYTHIA simulation containing 10\% photon exchange processes are
shown as stars.  The errors are statistical only.  A full detector 
simulation has been applied to the Monte Carlo event samples.
A comparison of the gap--fractions for data and Monte Carlo events 
in Figure~\ref{f:cgap2} reveals an excess in the fraction of gap events in 
the data over that expected for standard hard photoproduction processes.
Additionally, the data exhibit a two--component behaviour.  There is
an exponential fall at low values of $\Delta\eta$, but there is little
or no dependence of $f(\Delta\eta)$ on $\Delta\eta$ for $\Delta\eta > 3.2$.
We recall the definition of diffraction proposed in Sect.~\ref{s:DIFF}.
One is tempted to interpret the exponential fall of $f(\Delta\eta)$ as being
due to the production of gaps in non-diffractive processes.  Then the
flat component which dominates the rate of rapidity gap event production 
at large $\Delta\eta$ may be naturally interpreted as arising from a 
diffractive process.  However we must check first that this two--component
behaviour of $f(\Delta\eta)$ survives a full correction for detector
acceptance and smearing.  A detailed description of the correction method
and the assignment of systematic errors may be obtained 
elsewhere~\cite{Zrgbj,Laurel}.  

\begin{figure}[h]
\epsfxsize=12cm
\centering
\leavevmode
\epsfbox{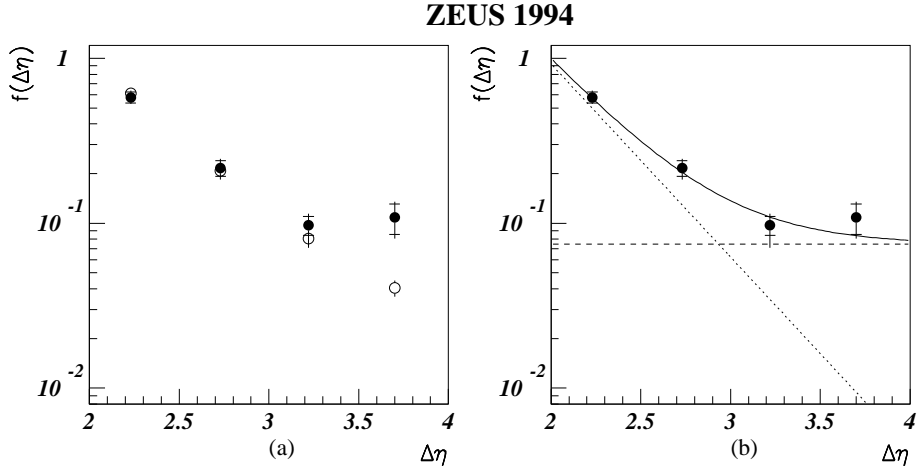}
\caption{\capstyle\label{f:cgap3}
        Corrected gap--fraction.  The data are shown as black dots.  The 
        inner error bar shows the statistical error and the outer error
        bar shows the systematic uncertainty combined in quadrature with
        the statistical error.  The open circles in (a) show the expectation
        from PYTHIA for standard hard photoproduction processes.  The 
        solid line in (b) shows the result of a fit to an exponential plus
        a constant dependence where the dotted and dashed lines show
        the exponential and constant terms respectively.}
\end{figure}

The measured gap--fraction, corrected for detector effects, is shown in 
Figure~\ref{f:cgap3} (black dots).  The statistical errors are shown by the
inner error bar and the systematic uncertainties combined in quadrature with
the statistical errors are indicated by the outer error bars.  (The data are
the same in Figures~\ref{f:cgap3}(a) and (b).)
Although there is some migration of events the overall detector corrections 
do not significantly affect the gap--fraction.

The corrected gap--fraction is compared with the prediction of the PYTHIA
Monte Carlo program for standard hard photoproduction processes in 
Figure~\ref{f:cgap3}(a).  (The PYTHIA prediction is shown by the open 
circles.) 
There is a significant discrepancy between the data and the prediction in
the bin corresponding to $\Delta\eta > 3.5$.  If we let the Monte Carlo
prediction represent our expectation for the behaviour of the gap--fraction
for non-diffractive processes then we can obtain an estimate of the
diffractive contribution to the data by subtracting the Monte Carlo
gap--fraction from the data gap--fraction.  We obtain $.07 \pm .03$.
Therefore we estimate that 7\% of the data are due to hard diffractive 
processes.

In Figure~\ref{f:cgap3}(b) a second method of estimating the contribution from
diffractive processes is illustrated.  Here we have made direct use of the
definition of diffraction quoted in Sect.~\ref{s:DIFF}.  We have performed
a two--parameter $\chi^2$ fit of the data to the sum of an exponential
term and a constant term, constraining the sum to equal 1 at 
$\Delta\eta = 2$.  (Below $\Delta\eta = 2$ the jet cones are overlapping in 
$\eta$.)  The diffractive
contribution which is the magnitude of the constant term
is thus obtained from all four of the measured data points.
It is $0.07 \pm 0.02(stat.) ^{+0.01} _{-0.02}(sys.)$ or again, 7\% of the
data are due to hard diffractive processes.

A caveat is in order.  Implicit in both methods of estimating the fraction
of diffractive processes in the data is the assumption that exactly 100\%
of hard diffractive scatterings will give rise to a rapidity gap.  In fact
this is considered to be an overestimate.  Interactions between the
$\gamma$ and $p$ spectator particles can occur which would fill in the
gap.  Therefore the result of $0.07 \pm 0.02(stat.) ^{+0.01} _{-0.02}(sys.)$
should be interpreted as a lower limit on the fraction of hard diffractive
processes present in the data.

The probability of no secondary interaction occurring has been
called the gap survival probability~\cite{bjorken}.  Estimates for 
the survival probability in 
$pp$ interactions range between 5\% and 30\%~\cite{bjorken,surv1,surv2}.
However for these $\gamma p$ collisions we expect the survival probability
to be higher due (in part) to the high
values of $x_{\gamma}$ of this data sample compared to typical values of
$x_p$ in a $pp$ data 
sample.\footnote{For instance, a typical event with two jets of 
$E_T^{jet} = 6$~GeV at $\Delta\eta$ = 3 at HERA would have $x_{\gamma} = 0.8$
while the corresponding event at the Tevatron with two jets of 
$E_T^{jet} = 30$~GeV and $\Delta\eta = 3$ would have $x_p = 0.09$.}
Therefore we do not consider the ZEUS
result to be incompatible with the D0 result, 
$0.0107 \pm 0.0010(stat.) ^{+0.0025}_{-0.0013}(sys.)$~\cite{GAPSD02}, and the
CDF result,  $0.0086 \pm 0.0012$~\cite{GAPSCDF}.

In summary, ZEUS has measured the fraction of dijet events which contain
a rapidity gap between the jets, $f(\Delta\eta)$.  
From a comparison of the uncorrected $f(\Delta\eta)$ with
that obtained from the PYTHIA simulation of hard 
photoproduction processes (with full detector simulation) we conclude that
the data are inconsistent with a completely non-diffractive production
mechanism.  From the behaviour of the fully corrected $f(\Delta\eta)$, 
we determine that the hard diffractive contribution to the dijet sample
is greater than $(7 \pm 3)\%$.  This value is obtained for two different
methods of estimating the non-diffractive contribution, {\em i)} 
letting the non-diffractive contribution be represented by the PYTHIA 
prediction for hard photoproduction processes and {\em ii)} 
obtaining the non-diffractive contribution directly from an 
exponential fit to the data.

\section{Forward Rapidity Gaps\label{s:FRG}}

\begin{figure}[h]
\epsfxsize=8cm
\centering
\leavevmode
\rotatebox{270}{\epsfbox{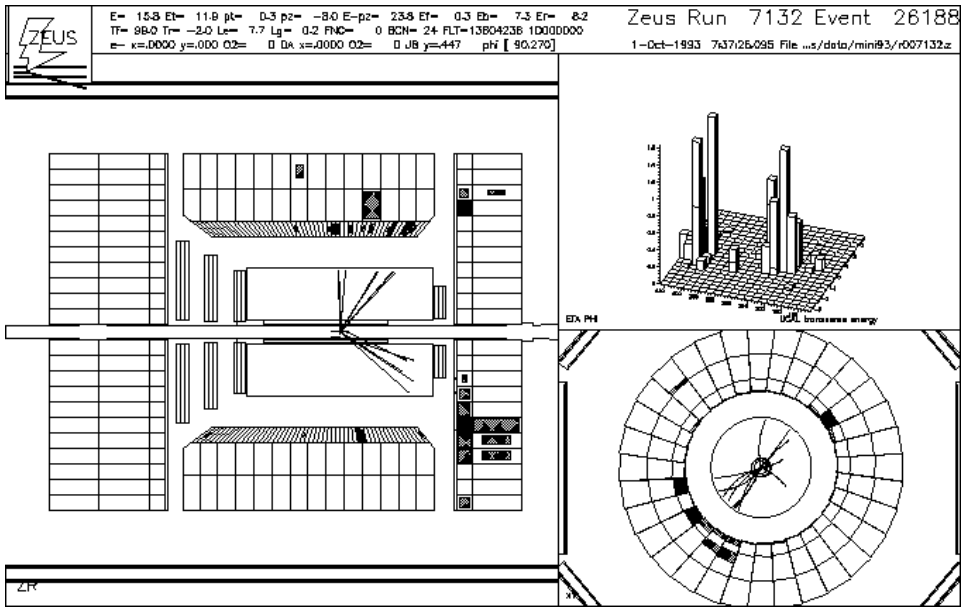}}
\caption{\capstyle\label{f:fgapevt}
        A hard photoproduction event with a foward gap.  The $z-R$ display
        of the ZEUS detector is shown on the left hand side.  In the
        upper right hand corner the $\eta$ and $\varphi$ coordinates of the
        calorimeter energy deposits are shown, weighted by their transverse
        energy.  The lower right hand view is the $x-y$ cross section.}
\end{figure}

The class of events which will be discussed in this section exhibits a 
rapidity gap extending to high values of $\eta$.  An example is shown in
Figure~\ref{f:fgapevt}.
There are two high transverse energy jets which are back to back in $\varphi$
and no scattered $e^-$ candidate.  This is a hard photoproduction
event.  However there is no energy in the forward direction around the
beam pipe which could be associated with the fragmentation products of the
proton remnant.  This is, therefore, a candidate diffractive hard scattering
event.

This analysis proceeds in a similar way to that described in the previous
section.  First the uncorrected data are compared to Monte Carlo generated
event samples which have been subjected to a full simulation of the ZEUS
detector.  In a second step the generated event samples are used to correct 
the data for the effects of the detector smearing and acceptance.  Again,
specific details of the analysis should be obtained from the 
publication~\cite{Zlowt3}.

\begin{figure}[h]
\epsfxsize=7cm
\centering
\leavevmode
\epsfbox{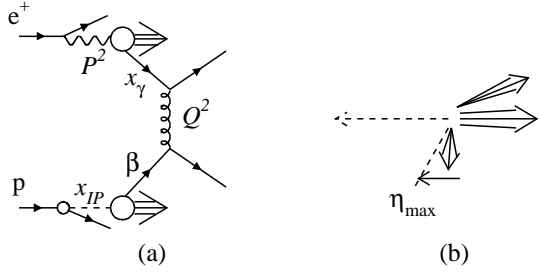}
\caption{\capstyle\label{f:lowt2}
        Kinematics of the diffractive hard photoproduction process.
        The meaning of the momentum fraction variables $x_{I\!\!P}$ and
        $\beta$ is illustrated in (a) where the fraction of the photon's
        momentum entering the hard subprocess, $x_{\gamma}$, the photon
        virtuality, $P^2$, and the squared momentum transfer which sets the
        energy scale of the hard subprocess,
        $Q^2$, are also indicated.  The pseudorapidity of the particle with
        the highest pseudorapidity is denoted $\eta_{max}$ as illustrated
        in (b).}
\end{figure}

The diffractive hard scattering process is understood to proceed as 
illustrated in Figure~\ref{f:lowt2}(a) where we have introduced two new momentum
fraction variables.  $x_{I\!\!P}$ represents the fraction of the proton's
momentum which is carried by the pomeron and $\beta$ represents the fraction
of the pomeron's momentum which is carried into the hard subprocess.  Of
course $x_{I\!\!P} \cdot \beta$ gives the familiar Bjorken--$x_p$ variable.
The other important variable for describing the diffractive hard 
photoproduction process is $\eta_{max}$.  $\eta_{max}$ is defined to be the 
pseudorapidity of the most forward going particle (measured using the 
calorimeter) which has energy exceeding 400~MeV.  The definition of 
$\eta_{max}$ is illustrated schematically in Figure~\ref{f:lowt2}(b).

\begin{figure}[h]
\epsfxsize=7cm
\centering
\leavevmode
\epsfbox{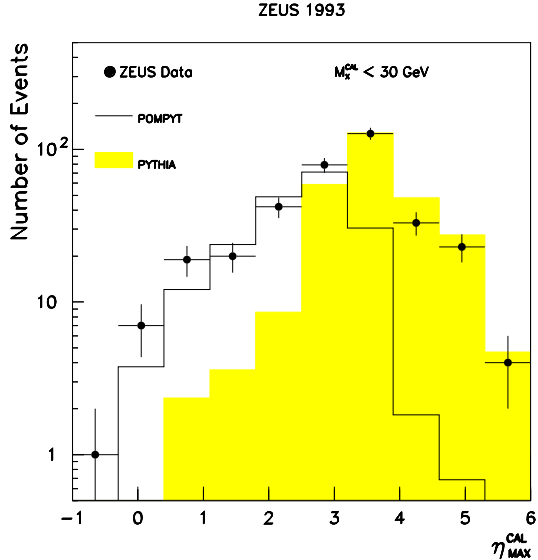}
\caption{\capstyle\label{f:fgap1}
        The $\eta_{max}$ distribution for a sample of hard photoproduction
        events.  The data are shown by black dots
        with error bars representing statistical errors only.  No corrections
        for detector effects have been made.  The shaded histogram shows
        the prediction from the PYTHIA standard hard photoproduction events.
        The open histogram shows the prediction from the POMPYT 
        simulation of $\gamma I\!\!P$ scattering where the $I\!\!P$ contains
        a hard gluon spectrum (see text).  The Monte Carlo event samples
        have been subjected to a full simulation of the detector acceptance
        and smearing.}
\end{figure}

The $\eta_{max}$ distribution for a sample of hard photoproduction events
is shown in Figure~\ref{f:fgap1}.  For this particular plot a subsample
of events is shown for which the total hadronic invariant mass, $M_X$, 
(measured using all of the energy deposits in the calorimeter) satisfies
$M_X < 30$~GeV.
The data are shown by black dots and are
not corrected for detector effects.  The errors shown are statistical only.
The data are peaked toward a value of $\eta_{max}$ which is close to the
edge of the calorimeter acceptance.  However there is a large contribution
from events with very low values of $\eta_{max}$, indicating the presence of
a forward rapidity gap.  Also shown in this figure are two Monte Carlo
predictions which include a full simulation of the ZEUS detector.  The
shaded histogram shows the PYTHIA prediction for standard hard photoproduction
processes.  It fails to describe the large rapidity gap events of the data
which occur at low values of $\eta_{max}$.  To describe these large rapidity
gap events we must introduce the Monte Carlo program POMPYT~\cite{POMPYT}, 
the prediction of which is shown by the open histogram.

POMPYT is a Monte Carlo implementation of the Ingelman--Schlein 
model~\cite{IngSchlein} which assumes that the hard photoproduction 
cross section $\sigma^{jet}_{\gamma p}$ factorizes in the following way.
\begin{equation}\label{e:fact}
   \sigma^{jet}_{\gamma p} = f_{I\!\!P / p}(x_{I\!\!P},t) \otimes
          f_{a / I\!\!P}(\beta, Q^2) \otimes \hat{\sigma}(\hat{s},Q^2).
\end{equation}
In words, the jet cross section, $\sigma^{jet}_{\gamma p}$, may be written
as the convolution of a term representing the flux of pomerons in the
proton, $f_{I\!\!P / p}(x_{I\!\!P},t)$, with a term describing the
flux of partons in the pomeron, $f_{a / I\!\!P}(\beta, Q^2)$, and with
the subprocess cross section, $\hat{\sigma}(\hat{s},Q^2)$.
The direct photoproduction subprocess cross section includes only the hard 
subprocesses, $\gamma q \rightarrow q g$ and $\gamma g \rightarrow q q$.
In resolved photoproduction it includes in addition to
the hard subprocesses $qq \rightarrow qq$, $qg \rightarrow qg$, etcetera,
the flux of partons in the photon, $f_{a / \gamma}(x_{\gamma},Q^2)$.
The hard subprocess cross sections are calculable in perturbative QCD and
some experimental information exists which constrains the 
$f_{a / \gamma}(x_{\gamma},Q^2)$.  Therefore $\hat{\sigma}(\hat{s},Q^2)$ is
a known input in Eqn.~\ref{e:fact}.  The pomeron flux factor,
$f_{I\!\!P / p}(x_{I\!\!P},t)$, may be
determined using Regge inspired fits to hadron--hadron data.  
The remaining unknown ingredient is the pomeron structure.  We neglect
the energy scale dependence of $f_{a / I\!\!P}(\beta, Q^2)$ and consider 
two extreme possibilities for its $\beta$ dependence.
The first, $\beta f_{a / I\!\!P}(\beta) = 6 \beta (1 - \beta)$, yields
a mean $I\!\!P$ momentum fraction of $\langle \beta \rangle = 1 / 2$ and
is therefore known as the hard parton density.
The second, $\beta f_{a / I\!\!P}(\beta) = 6 (1 - \beta) ^ 5$, has
$\langle \beta \rangle = 1 / 7$ and is called the soft parton density.
Finally, it is not clear whether there should be a momentum sum rule
for the pomeron, that is, whether 
$\Sigma_{I\!\!P} \equiv \int_0^1 d\beta \sum_a \beta f_{a / I\!\!P}(\beta)$
must equal 1 or not.

The open histogram in Figure~\ref{f:fgap1} shows the POMPYT prediction for
a $I\!\!P$ consisting entirely of gluons with the hard momentum spectrum.
A fairly satisfactory description of $\eta_{max}$ may be achieved.  In 
addition
the POMPYT prediction is able to describe the $M_X$ distribution, and
the distribution of the photon proton centre--of--mass energies, 
$W_{\gamma p}$, for rapidity gap events with $\eta_{max} < 1.8$.
(The results are similar for a $I\!\!P$ composed entirely of hard quarks.)
For this reason we say that the data are consistent with containing a
contribution from diffractive hard photoproduction processes.

\begin{figure}[h]
\epsfxsize=7cm
\centering
\leavevmode
\epsfbox{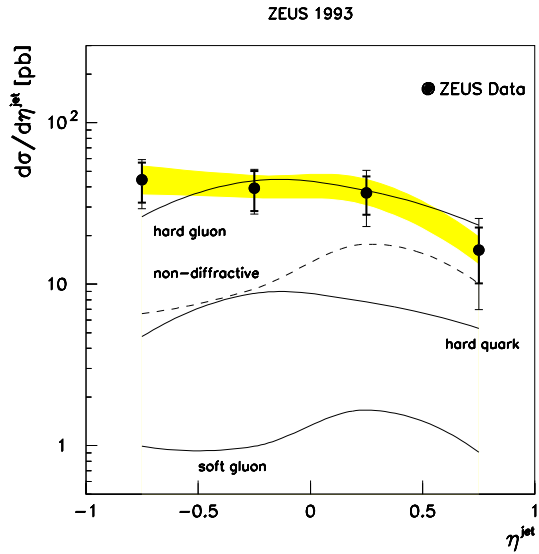}
\caption{\capstyle\label{f:fgap2}
        Cross section $d \sigma / d \eta^{jet}$ for photoproduction of jets 
        with $E_T^{jet} > 8$~GeV in events with $\eta_{max} < 1.8$.  The
        inner error bars show the statistical errors and the outer error bars
        the systematic uncertainty added in quadrature --- excluding the 
        systematic uncertainty due to the calorimeter energy scale which is 
        shown by
        the shaded band.  The PYTHIA prediction for standard hard 
        photoproduction processes is shown by the dashed line.  The POMPYT
        predictions $d\sigma^{\mbox{\scriptsize\it hard gluon}}/d\eta^{jet}$,
        $d\sigma^{\mbox{\scriptsize\it hard quark}}/d\eta^{jet}$ and
        $d\sigma^{\mbox{\scriptsize\it soft gluon}}/d\eta^{jet}$
        for diffractive hard processes with different
        parton distribution functions and $\Sigma_{I\!\!P} = 1$ are shown by 
        the solid lines.}
\end{figure}

In the second step of the analysis we correct the data for all effects of
detector acceptance and smearing.  We present in Figure~\ref{f:fgap2} the
$ep$ cross section for photoproduction of jets of $E_T^{jet} > 8$~GeV
as a function of the jet pseudorapidity.
This cross section is for events which have a rapidity gap characterized 
by $\eta_{max} < 1.8$.
The PYTHIA prediction for this cross section for non-diffractive processes
is shown by the dashed line.  It is too low in overall magnitude to describe
the data as well as being disfavoured in shape.  The POMPYT predictions, 
$d\sigma^{\mbox{\scriptsize\it hard gluon}}/d\eta^{jet}$,
$d\sigma^{\mbox{\scriptsize\it hard quark}}/d\eta^{jet}$ and
$d\sigma^{\mbox{\scriptsize\it soft gluon}}/d\eta^{jet}$ 
for the hard gluon, the hard quark and the soft gluon pomeron parton 
densities respectively where $\Sigma_{I\!\!P} = 1$ are shown by the 
solid curves.
The soft parton density very rarely gives rise to sufficient momentum transfer
to produce two $E_T^{jet} > 8$~GeV jets and so
$d\sigma^{\mbox{\scriptsize\it soft gluon}}/d\eta^{jet}$ 
lies far below the cross sections for the hard parton densities 
in overall normalization.
$d\sigma^{\mbox{\scriptsize\it soft gluon}}/d\eta^{jet}$ is inconsistent
with the data in overall magnitude as well as being disfavoured in shape.
We do not consider soft parton densities further.
$d\sigma^{\mbox{\scriptsize\it hard quark}}/d\eta^{jet}$
is consistent with the data in shape but too small in 
magnitude.  
$d\sigma^{\mbox{\scriptsize\it hard gluon}}/d\eta^{jet}$
is capable of describing both the shape
and magnitude of the measured cross section.
Note, however, that the non-diffractive contribution to the data has not been 
subtracted, nor has the double dissociation contribution.

\begin{figure}[h]
\epsfxsize=7cm
\centering
\leavevmode
\epsfbox{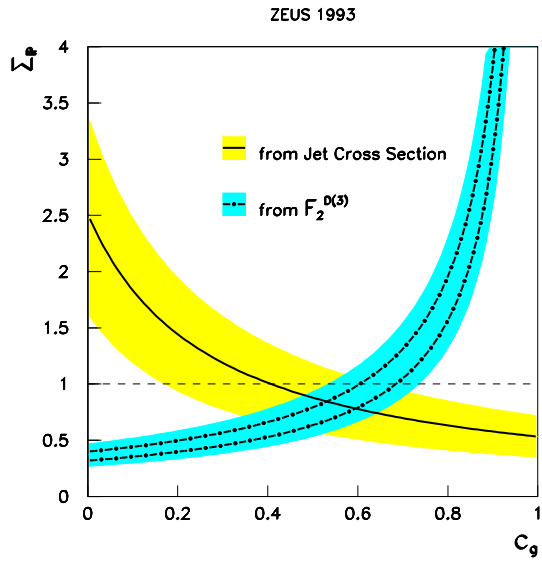}
\caption{\capstyle\label{f:fgap3}
        Allowed regions of the $\Sigma_{I\!\!P} - c_g$ plane.  The solid
        line and its shaded band of uncertainty show the constraint from 
        the measurement of $d\sigma / d\eta^{jet}$.  The 
        dash--dotted lines and their shaded band of uncertainty show the
        constraint imposed from the measurement of $F_2^{D(3)}$.  (The upper
        dash--dotted curve is for two quark flavours and the lower 
        dash--dotted curve is for three quark flavours.)}
\end{figure}

In the final stage of the analysis the non-diffractive contribution was
subtracted from the data using the PYTHIA prediction (which has been
shown to provide a good description of inclusive jet cross sections
in photoproduction~\cite{Zincj93}).  
A contribution
of $(15 \pm 10)$\% due to double dissociation processes was also subtracted.
Then the assumption that $\Sigma_{I\!\!P} = 1$ was relaxed.  The $I\!\!P$
was assumed to be composed of a fraction $c_g$ of hard gluons and a fraction
$1-c_g$ of hard quarks.
Then for various values of $c_g$ the expression
$\Sigma_{I\!\!P} \cdot 
      [ c_g \cdot d\sigma^{\mbox{\scriptsize\it hard gluon}}/d\eta^{jet} 
  + (1-c_g) \cdot d\sigma^{\mbox{\scriptsize\it hard quark}}/d\eta^{jet} ] $ 
was fit to the measured $d\sigma / d\eta^{jet}$ distribution to obtain
$\Sigma_{I\!\!P}$.  (The POMPYT predictions for 
$d\sigma^{\mbox{\scriptsize\it hard gluon}}/d\eta^{jet}$ and
$d\sigma^{\mbox{\scriptsize\it hard quark}}/d\eta^{jet}$ were used in the
fit.)
The result of this series of fits is shown in Figure~\ref{f:fgap3} by the
solid line where the statistical uncertainty of the fit is indicated by the 
shaded band.
We find, for instance, that the data are do not favour a $I\!\!P$ which
consists exclusively of hard gluons and simultaneously satisfies
the momentum sum rule, $\Sigma_{I\!\!P} = 1$.  (See~\cite{Zlowt3}, however,
for a discussion of additional theoretical systematic uncertainties.)

Results from studies of diffractive hard electroproduction have 
been expressed in terms of the diffractive structure function,
$F_2^{D(3)}(\beta,Q^2,x_{I\!\!P})$~\cite{Hlowt2,Zlowt2}.  The expression of
factorization is then, 
$F_2^{D(3)}(\beta,Q^2,x_{I\!\!P}) = f_{I\!\!P / p}(x_{I\!\!P}) \cdot 
                                          F_2^{I\!\!P}(\beta,Q^2)$.
Integrating this over $x_{I\!\!P}$ and
$\beta$ and then subtracting 
the integral over the pomeron flux
thus gives the sum of the momenta of all of the quarks in the pomeron,
$\Sigma_{I\!\!P} \cdot (1 - c_g)$.
The ZEUS measurement~\cite{Zlowt2}, 
$\Sigma_{I\!\!P} \cdot (1 - c_g) = 0.32 \pm 0.05$, is shown in 
Figure~\ref{f:fgap3} by the lower dot--dashed line.  This is the result for
two flavours of quark in the $I\!\!P$.  The result for three flavours
of quark is $\Sigma_{I\!\!P} \cdot (1 - c_g) = 0.40 \pm 0.07$, the upper
dot--dashed line in Figure~\ref{f:fgap3}.  The dark--shaded band shows the
additional measurement uncertainty.

Assuming that the pomeron flux is the same in the measurement of 
$F_2^{D(3)}$ and of $d\sigma^{jet}/d\eta^{jet}(\eta_{max} < 1.8)$ 
one can combine the two analyses to determine the allowed ranges,
$0.5 < \Sigma_{I\!\!P} < 1.1$ and $0.35 < c_g < 0.7$.  However the
$\Sigma_{I\!\!P}$ range is affected by additional uncertainties in the
normalization of the pomeron flux factor.  Taking into account all remaining
systematic uncertainties of the measurements we find $0.3 < c_g < 0.8$.
This measurement is independent of the pomeron flux and of the total momentum
carried by partons in the pomeron.

In summary, the distributions of $\eta_{max}$, $M_X$ and $W_{\gamma p}$
indicate that the events with a forward rapidity gap are consistent with
a diffractive hard scattering via exchange of a low--$t$ pomeron.
The $ep$ cross section $d\sigma/d\eta^{jet}$ for the photoproduction of
jets of $E_T^{jet} > 8$~GeV in large rapidity gap events ($\eta_{max} < 1.8$)
has been measured and is significantly larger than the cross section due to 
non-diffractive processes.  A comparison of the $d\sigma/d\eta^{jet}$ 
measurement in 
photoproduction with the measurement of $F_2^{D(3)}$ in electroproduction
indicates that 30\% to 80\% of the momentum of the pomeron which is due
to partons is carried by hard gluons.

\section{Conclusions and Outlook}
Evidence is being accumulated which indicates that there is a 
strongly interacting colour
singlet object which can mediate high--$t$ interactions and which also 
contributes through its partonic content to low--$t$ interactions.
Further work to extrapolate the diffractive cross section to intermediate $t$ 
ranges by determining its $t$ dependence may bring about a confrontation of 
the experimental results in these complementary regimes.  The Tevatron and 
HERA results pertaining to hard diffractive scattering cannot be directly 
compared at the moment, due to a lack of understanding of the gap survival
probabilities.  One possible route to achieve a more stringent comparison
of the Tevatron and HERA data may be for the Tevatron experiments to try
to measure the diffractive contribution to their data in a regime where
the survival probability is expected to be high, i.e., for a sample with
very high $x_p$.  (The HERA experiments cannot do the converse and go
to very low $x_{\gamma}$ while remaining in the regime of applicability
of perturbative QCD.)  The Tevatron constraint on the 
$\Sigma_{I\!\!P} - c_g$ plane from measurements of diffractive hard 
scattering is only barely consistent with the HERA constraint
at present.  (They find, for instance, that $\Sigma_{I\!\!P}$ must be
less than 0.5 if $c_g \sim 0.5$~\cite{dino96}.)  
We look forward to an exciting comparison in
the near future.  As for the confrontation between experiment and theory, 
neither are presently precise enough for any strong statements
to be made and much work remains to be done.

\section{Acknowledgements}
It is a pleasure to acknowledge assistance from Tony Doyle, Claudia Glasman,
Dino Goulianos and Brent May.
%
%\bibliography{/zow/users/sinclair/biblio/csin95}

%
\end{document}